\documentclass[letterpaper, 10 pt, conference]{ieeeconf}
\IEEEoverridecommandlockouts
\makeatletter
\let\NAT@parse\undefined
\makeatother
\usepackage{cite}
\usepackage{amsmath,amssymb,amsfonts}
\usepackage{algorithmic}
\usepackage{graphicx}
\usepackage{textcomp}
\usepackage{xcolor}
\usepackage{hyperref}
 \usepackage[aboveskip=1pt]{subcaption}
\usepackage[skip=0.75pt]{caption}
\usepackage{multirow}
\usepackage{hyperref}
\usepackage{multicol}
\usepackage{booktabs}
\usepackage{graphicx}
\usepackage{color}
\usepackage{url}
\def\BibTeX{{\rm B\kern-.05em{\sc i\kern-.025em b}\kern-.08em
    T\kern-.1667em\lower.7ex\hbox{E}\kern-.125emX}}
\begin{document}

\title{\bf Expert-Agnostic Ultrasound Image Quality Assessment using \\Deep Variational Clustering\\
}
\author{Deepak Raina$^{12*}$, Dimitrios Ntentia$^{23}$, SH Chandrashekhara$^4$, Richard Voyles$^2$, Subir Kumar Saha$^1$\\\vspace{-0.5cm}
\thanks{This work was supported in part by SERB (India) - OVDF Award No. SB/S9/Z-03/2017-VIII; PMRF - IIT Delhi under Ref. F.No.35-5/2017-TS.I:PMRF; Daniel C. Lewis Professorship; Berea College, Kentucky, USA and PU-IUPUI Collaborative Seed Grant.}
\thanks{$^{1}$Indian Institute of Technology (IIT), Delhi, India (\{deepak.raina, saha\}@mech.iitd.ac.in); $^{2}$Purdue University (PU), Indiana, USA (\{draina, ntentiad, rvoyles\}@purdue.edu) $^{3}$Berea College, Kentucky, USA (ntentiad@berea.edu); $^{4}$All India Institute of Medical Sciences (AIIMS), Delhi, India (drchandruradioaiims@gmail.com).}
\thanks{$^{*}$Corresponding author is Deepak Raina}
}


\maketitle
\begin{abstract}
Ultrasound imaging is a commonly used modality for several diagnostic and therapeutic procedures. However, the diagnosis by ultrasound relies heavily on the quality of images assessed manually by sonographers, which diminishes the objectivity of the diagnosis and makes it operator-dependent. The supervised learning-based methods for automated quality assessment require manually annotated datasets, which are highly labour-intensive to acquire. These ultrasound images are low in quality and suffer from noisy annotations caused by inter-observer perceptual variations, which hampers learning efficiency. We propose an UnSupervised UltraSound image Quality assessment Network, US2QNet, that eliminates the burden and uncertainty of manual annotations. US2QNet uses the variational autoencoder embedded with the three modules, pre-processing, clustering and post-processing, to jointly enhance, extract, cluster and visualize the quality feature representation of ultrasound images. The pre-processing module uses filtering of images to point the network's attention towards salient quality features, rather than getting distracted by noise. Post-processing is proposed for visualizing the clusters of feature representations in 2D space. We validated the proposed framework for quality assessment of the urinary bladder ultrasound images. The proposed framework achieved $\mathbf{78\%}$ accuracy and superior performance to state-of-the-art clustering methods. The project page with source codes is available at \href{https://sites.google.com/view/us2qnet}{\color{blue}\textit{{https://sites.google.com/view/us2qnet}}}.
\end{abstract}

\section{Introduction}
UltraSound (US) is the most frequently employed medical imaging modality in clinical practice due to its real-time feedback, low-cost, portability, and non-ionizing nature. It is useful for diagnosis, pre- and post-operative assessment, and surgical interventions. However, the diagnosis by ultrasound depends a lot on the image quality, which is determined manually by sonographers during image acquisition \cite{welleweerd2020automated}. Ultrasound images are quite difficult to interpret due to noise, shadows, poor contrast, and other sensors and motion artifacts \cite{noble2016reflections}. The blurred boundaries and presence of multiple anatomical structures further increase the difficulty in ensuring quality during acquisition. 
\begin{figure}[ht]
    \centering
	\includegraphics[trim=0.1cm 10.5cm 0.1cm 3.1cm,clip,width=\linewidth]{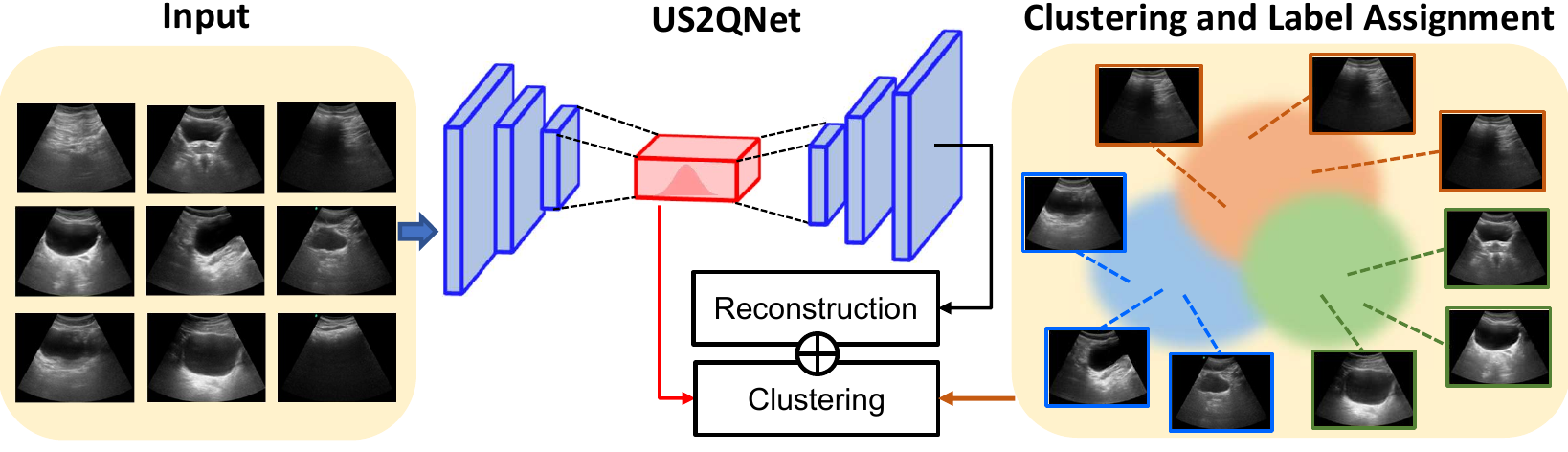}
	\caption{A preview of the framework for US-IQA using deep variational clustering. The circular clusters represent the ultrasound image classification based on their quality.}
    \label{fig:teaser} 
\end{figure}
Thus, the requirement of minimal manual effort and ensuring consistent quality among novice sonographers during image acquisition has prompted the research community to use automated methods for US Image Quality Assessment (US-IQA).


The supervised learning-based methods using a deep Convolutional Neural Network (CNN) have shown promising results in medical imaging analysis, including segmentation and classification using curated and annotated large-scale datasets \cite{shen2017deep}. However, the manual data annotation process is quite difficult for US image quality, as the US images with the same quality have significant differences and images with significant differences in quality appear similar. Moreover, the image quality annotation is noisy due to inter- and intra-observer perceptual variations \cite{wang2021annotation}. Thus, the annotation process is often labour-intensive, time-consuming and requires the participation of more than one expert radiologists, thereby making supervised learning quite challenging for automated US-IQA. 

In this work, we propose an \textbf{U}n\textbf{S}upervised learning-based \textbf{U}ltra\textbf{S}ound image \textbf{Q}uality assessment \textbf{Net}work, termed as \textbf{US2QNet}, as shown in Fig. \ref{fig:teaser}. The objective of the work is to address the complexity and uncertainty in manual annotation of US image quality by expert radiologists for supervised learning. The \textit{key contributions} of framework are as follows: 
\begin{enumerate}
    \item We proposed the first deep clustering framework for US-IQA, which uses a variational autoencoder in conjunction with pre-processing, clustering and post-processing modules. The framework will jointly enhance, extract, cluster and visualize the quality feature representation of US images.
    \item We introduced US imaging-specific pre- and post-processing modules in deep clustering. Pre-processing uses fuzzy filtering to direct the network's attention towards the critical quality features. The post-processing module generates the visualization of high-dimensional quality feature representations in 2D space and provides an effective way to do hyperparameter tuning.
    \item The proposed framework was validated and compared to the state-of-the-art (SOTA) methods for urinary bladder US images. The results revealed that the proposed method is effective for unsupervised US-IQA and outperformed the SOTA methods.
\end{enumerate}  
The unsupervised deep clustering approach strategy has previously been investigated mostly for natural images \cite{10.1007/978-3-319-70096-0_39, caron2018deep, caron2019unsupervised}, however, to the best of the author's knowledge, this is the first attempt to use this approach for the challenging task of classifying the US images based on quality.
\subsection{Related work} \label{sec:literature}
\noindent
\textbf{Supervised learning for US-IQA:}
The availability of annotated medical images has motivated the researchers to use supervised training of CNNs for automated analysis of X-ray, Magnetic Resonance Imaging (MRI), and US imaging \cite{shen2017deep}. For US-IQA, Wu \textit{et al.} \cite{wu2017fuiqa} proposed the two deep CNNs for jointly finding the region of interest and assessing the fetal ultrasound image quality. Lin \textit{et al.} \cite{lin2018quality, lin2019multi} used a multi-task faster regional CNN architecture to first localize the six key anatomical structures of the fetal head and then assigned the quality score based upon their clear visibility. However, they employed prior clinical knowledge to determine the relative position of anatomical structures. In addition, the fetal US datasets in \cite{wu2017fuiqa, lin2018quality, lin2019multi} require substantial clinical expertise for labeling the region of interest and anatomical structures with comparable appearances such as stomach bubble, umbilical vein, choroid plexus, and others. Moreover, the performance of supervised IQA models is impacted by annotation noise due to inter- and intra-observer perceptual variations \cite{wang2021annotation}.
\\
\\
\noindent
\textbf{Unsupervised learning for medical imaging:}
In order to overcome the challenges of labour-intensive annotation in supervised learning, different techniques of unsupervised learning have been explored for medical imaging analysis \cite{raza2021tour}. These techniques used traditional clustering methods \cite{zhang2021cnn}, modern methods like Autoencoder (AE), generative networks \cite{wang2022sfgan}, and Deep Clustering \cite{zhou2022comprehensive}. AE is one of the most effective algorithms for unsupervised feature extraction and compression to low-dimensional latent space while minimizing the reconstruction error between the encoded input and decoded output. They have been mostly used for denoising and anomaly detection in medical images \cite{xu2021review}. In addition, Generative networks like Variational AE (VAE) have also shown great potential for medical image analysis \cite{wei2020recent}. They use variational loss during training and enforce a predefined distribution in the latent space. A work by Nesovic \textit{et al.} \cite{nesovic2021ultrasound} attempted to address the US-IQA using AE combined with Random Forest Classifier (RFC). However, RFC required noise-specific hand-crafted features for quality classification, which are challenging to design for noisy US images \cite{noble2016reflections}. 
\\
\\
\noindent
\textbf{Deep clustering for image classification:}
Clustering is another powerful unsupervised method to detect patterns in non-labelled dataset. There exist several classical clustering algorithms like $k-$means, Principal Component Analysis (PCA),  Density-Based Spatial Clustering of Applications with Noise (DBSCAN) \cite{verma2012comparative}, Spectral clustering \cite{xu2012spectral}, t-distributed Stochastic Neighbor Embedding (t-SNE) \cite{van2008visualizing}, Uniform Manifold Approximation and Projection (UMAP) \cite{ghojogh2021uniform}. However, they can't handle high-dimensional data.
In recent years, deep clustering has gained the researcher's attention for unsupervised image analysis \cite{10.1007/978-3-319-70096-0_39,caron2018deep,caron2019unsupervised}. Deep clustering uses classical clustering methods in conjunction with AEs and VAEs to learn the clusterable representation in latent space. This has been used for the categorization of handwritten digits and face images \cite{song2013auto}, natural image datasets \cite{prasad2020variational}, Magnetic Resonance Imaging \cite{soleymani2021deep, maicas2020unsupervised} and Computed Tomography images \cite{moriya2018unsupervised}. Yu \textit{et al.} \cite{yu2022feature} proposed deep clustering for classifying thyroid cancer in US images.  Notably, the potential of deep clustering has not yet been uncovered for US-IQA, which is quite challenging to analyze due to the noise, artifacts, varying viewpoints, low inter-class and huge intra-class variability \cite{noble2016reflections}.
\section{Methodology}
\begin{figure*}[ht]
    \centering
	\includegraphics[trim=1cm 2cm 0.8cm 5.2cm,clip,width=\linewidth]{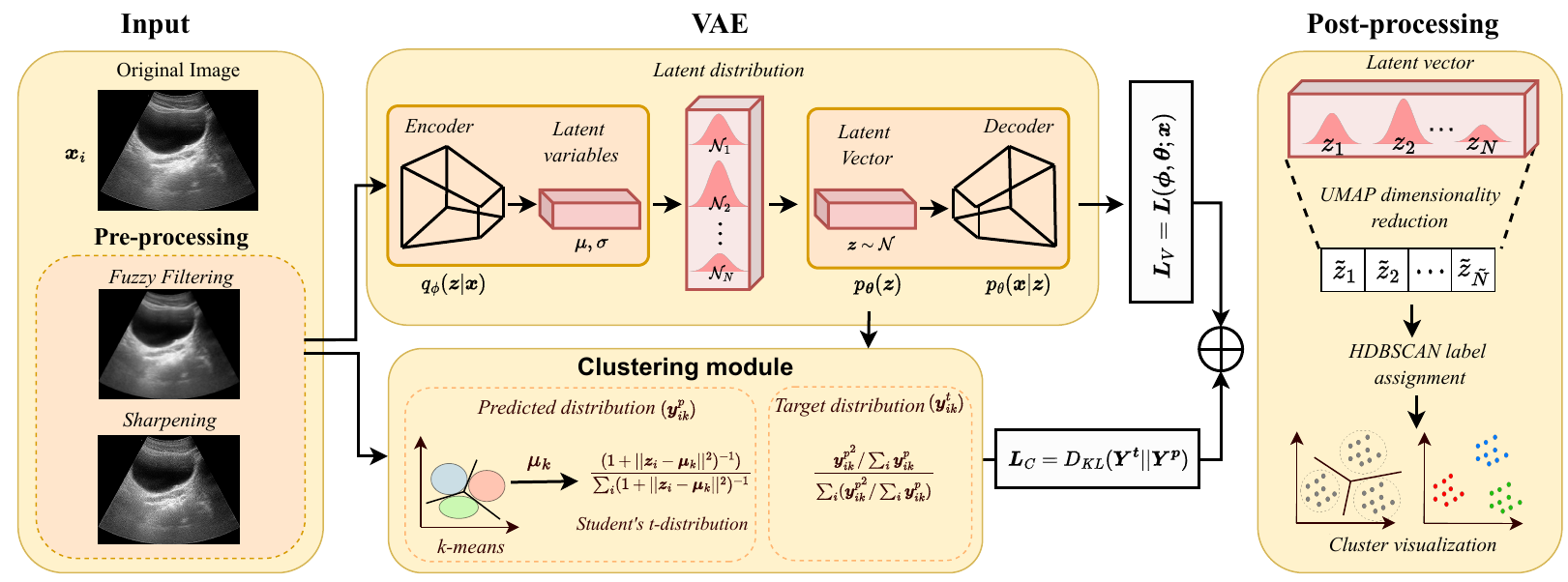}
	\caption{Overview of the US2QNet architecture for ultrasound image quality assessment using deep variational clustering.}
    \label{fig:arch} 
\end{figure*}
Fig. \ref{fig:arch} depicts the overview of the proposed framework. The framework is executed in three stages, which include: (1) Training the VAE on the pre-processed dataset using the reconstruction function as the loss function; (2) Fine-tuning the VAE by jointly optimizing the reconstruction loss and clustering loss for achieving clusterable quality feature representation of US images; and (3) Dimensionality reduction and visualization of learned feature representation using UMAP. Later, we utilized HDBSCAN for assigning labels to the clusters corresponding to the US image quality.
\subsection{Variaional Autoencoder} \label{sec:vae}
VAE is the generative variant of AE, as it enforces the encoder of AE to generate features representing a normal distribution $\mathcal{N}(\boldsymbol{\mu}, \boldsymbol{\sigma})$ in the latent space. VAEs have shown success in multi-class clustering due to their generative representation capability.
The VAE introduces and approximates a latent variable $\boldsymbol{z}$ that follows a prior distribution $p(\boldsymbol{z})$. The model then infers a posterior distribution $q_{\boldsymbol{\phi}}(\boldsymbol{z}|\boldsymbol{x})$, and an output of $p_{\boldsymbol{\theta}}(\boldsymbol{x}|\boldsymbol{z})$, with encoder parameterized by $\boldsymbol{\phi}$ and a decoder parameterized by $\boldsymbol{\theta}$. Therefore, the objective of variational inference is to precisely compute $p_{\boldsymbol{\theta}}(\boldsymbol{x}|\boldsymbol{z})$ by inferring the latent distribution from the observed dataset. 
During training, a latent vector $\boldsymbol{z}$ is sampled from $p_{\boldsymbol{\theta}}(\boldsymbol{z})$, and the decoder produces the parameters of $p_{\boldsymbol{\theta}}(\boldsymbol{x}|\boldsymbol{z})$, which we may use to sample an output vector $\boldsymbol{x}$. The following equation formulates the loss function of VAE as:
\begin{multline}
    \boldsymbol{L}(\boldsymbol{\phi}, \boldsymbol{\theta}; \boldsymbol{x}) = \mathbb{E}_{\boldsymbol{z}\sim q_{\boldsymbol{\phi}}(\boldsymbol{z}|\boldsymbol{x})} [\log p_{\boldsymbol{\theta}}(\boldsymbol{x}|\boldsymbol{z})] \\ + D_{KL}(q_{\boldsymbol{\phi}}(\boldsymbol{z}|\boldsymbol{x})||p_{\boldsymbol{\theta}}(\boldsymbol{z}))
\label{eq:loss_vae}
\end{multline}
where the first term is reconstruction error between observed data and data decoded from the latent vector and the second term is Kullback-Leibler ($KL$) divergence between the distribution of observed data $q_{\boldsymbol{\phi}}(\boldsymbol{z}|\boldsymbol{x})$ and a certain distribution $p_{\boldsymbol{\theta}}(\boldsymbol{z})$. 
The prior $p_{\boldsymbol{\theta}}(\boldsymbol{z})$ in VAE is often modeled by a single multivariate Gaussian in VAEs \cite{9207523}. Lin \textit{et al.} \cite{Lin_2019} proposed using a mixture of Gaussian as prior for unsupervised classification of the CIFAR Dataset and achieved remarkable results. Inspired by their work, we have used a mixture of Gaussian as prior $p_{\boldsymbol{\theta}}(\boldsymbol{z})$ for the latent space distribution representation to accurately describe the structure of each cluster using a specific Gaussian.
\\
\\
\textbf{Pretraining of VAE:} During pre-training, we used $\boldsymbol{L}(\boldsymbol{\phi}, \boldsymbol{\theta}; \boldsymbol{x})$ as the loss function (denoted as $\boldsymbol{L}_V$) to optimize the set of parameters $\boldsymbol{\theta}^*$ of VAE as
\begin{equation}
    \boldsymbol{\theta}^* = arg \max_{\boldsymbol{\theta}} \boldsymbol{L}(\boldsymbol{\phi}, \boldsymbol{\theta}; \boldsymbol{x})
\end{equation}
\subsection{Deep variational clustering} \label{sec:clustering}
Several algorithms have been proposed for deep clustering as described in Section \ref{sec:literature}. Among them, Deep Embedded Clustering (DEC) \cite{10.1007/978-3-319-70096-0_39} is one of the most representative methods of deep clustering, which proposed joint optimization of feature representation and clustering. In our framework, we used VAE, in contrast to AE in \cite{10.1007/978-3-319-70096-0_39} and embedded the DEC module between the encoder and decoder. We used the acronym DEC-VAE in the paper for this clustering approach. The joint loss function is then given by the sum of VAE loss ($\boldsymbol{L}_V$) and clustering loss and is formulated as follows:
\begin{equation}
\boldsymbol{L}=\boldsymbol{L}_{V}+\gamma \boldsymbol{L}_{C}
\label{loss function}
\end{equation}
where $\boldsymbol{L}_{C}$ is the clustering loss and $\gamma$ is a coefficient used to control the degree of distortion in the latent space.
\\
\\
\textbf{Clustering loss:}
The clustering module works like a validation set, where in each epoch, we used an auxiliary target distribution $\boldsymbol{Y^t}$ to iteratively correct the distribution of predicted label assignments $\boldsymbol{Y}^p$. To define clustering loss, we used the $KL$ divergence to decrease the difference between the $\boldsymbol{Y^t}$ and $\boldsymbol{Y^p}$ distribution as:
\begin{equation} \label{eq:lc}
    \boldsymbol{L}_{C}=D_{KL}(\boldsymbol{Y^t}||\boldsymbol{Y^p})= \sum_{i}^{N}\sum_{j}^{K}\boldsymbol{y}^t_{ik} \log \frac{\boldsymbol{y}^t_{ik}}{\boldsymbol{y}^p_{ik}}
\end{equation}
where $\boldsymbol{y}^p_{ik}$ denotes the probability of assigning cluster $\mu_k$ to latent feature $\boldsymbol{z}_i$ and $\boldsymbol{y}^t_{ik}$ is the auxiliary target distribution. The $\boldsymbol{y}^p_{ik}$ is calculated using Student's t-distribution as:
\begin{equation}
    \boldsymbol{y}^p_{ik}=\frac{(1+||\boldsymbol{z}_{i}-\boldsymbol{\mu}_{k}||^{2})^{-1}}{\sum_{i}(1+||\boldsymbol{z}_{i}-\boldsymbol{\mu}_{k}||^{2})^{-1}}
\end{equation}
where $\boldsymbol{\mu}_k$ is the initial cluster centroids and calculated by performing standard $k-$means clustering.  The target distribution $\boldsymbol{y}^t_{ik}$ is defined by the following equation:
\begin{equation}
    \boldsymbol{y}^t_{ik}=\frac{\boldsymbol{y}_{ik}^{p^2}/ \sum_{i}\boldsymbol{y}^p_{ik}} {\sum_{i}(\boldsymbol{y}_{ik}^{p^2}/\sum_{i}\boldsymbol{y}^p_{ik})}
\end{equation}
The clustering loss in eq. \eqref{eq:lc} is used to correct the weights,  representing cluster centers $\boldsymbol{\mu}_{k}^K$, and pivot the image latent feature $\boldsymbol{z}_i$ from a generative image representation to a representation focused on discriminative quality features in the latent space distribution. We have used three clusters ($K=3$) in our dataset representing three levels of US image quality.
\subsection{Post-processing}
The post-processing module is used to visualize the feature representation in 2D space, which provides an effective way to analyze the model's accuracy. The visualization of clusters also played an important role in the selection and iterative correction of the network hyper-parameters. After fine-tuning the network weights, the latent space feature vector is sampled from learned feature distribution as $[\boldsymbol{z}_1 \sim \mathcal{N}_1,\cdots,\boldsymbol{z}_N \sim \mathcal{N}_N]$. Then, the latent space feature vector of dimension $N$ is transformed to lower-dimensional space $\tilde{N}$ as $[\tilde{\boldsymbol{z}}_1,\cdots,\tilde{\boldsymbol{z}}_{\tilde{N}}]$ for efficient clustering of features. Several algorithms exist for dimensionality reduction and clustering analysis like PCA, $k-$means, t-SNE, UMAP, and HDBSCAN, as discussed in Section \ref{sec:literature}. Our exhaustive study on different clustering approaches revealed that UMAP for dimensionality reduction followed by HDBSCAN for label assignment outperformed all other combinations. The UMAP is chosen to ensure the efficient learning of manifold structure from a high-dimensional latent feature representation. In addition, UMAP returns reproducible results which is not the case in other dimensionality reduction methods like PCA. The capability of HDBSCAN to preserve both global and local density clusters while effectively separating overlapping data makes it a desirable option for label assignment \cite{blanco2022strategies}. 
\section{Dataset} \label{sec:dataset}
\begin{figure}[ht]
    \centering
     \includegraphics[trim=0cm 8cm 11cm 0cm,clip,width=\linewidth]{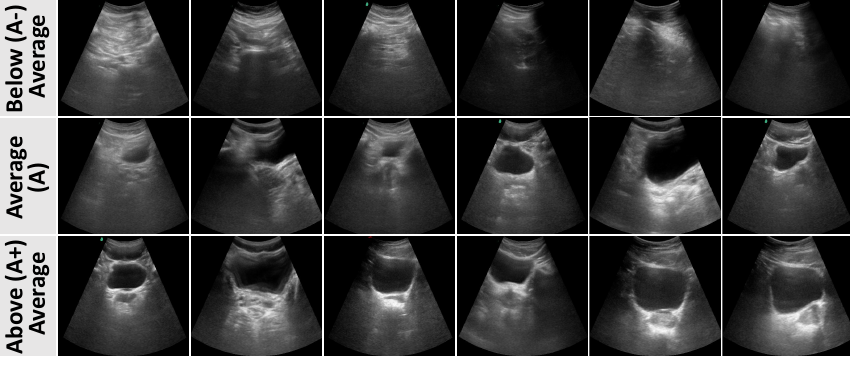}
    \caption{Sample test set images with their quality labels mentioned on the extreme left of each quality set.}
    \label{fig:dataset}
\end{figure}

\noindent
\textbf{Collection:}
The dataset consists of US images collected during the feasibility study of our in-house developed Telerobotic Ultrasound (TR-US) system \cite{tr-us, chandrashekhara2022robotic} at All India Institute of Medical Sciences (AIIMS), New Delhi (a public hospital and medical research university in India). The AIIMS Ethics Committee approved the study (Ref. No. IEC-855/04.09.2020,RP-16/2020). The study was conducted between October $2021$ and December $2021$. Before the scanning operation, informed written consent was obtained from the volunteers, and their privacy was protected by anonymizing their identifiers in the image. The dataset consists of male urinary bladder US images, which is used for discriminating the bladder shape, and identifying diverticula, stones, malignant tumours, and free fluid (blood) during trauma. The dataset is challenging to analyze, having spurious textures and inter- and intra-subject variations.  
\\
\\
\textbf{Pre-processing:}
The deep learning models are quite sensitive to noise in images, therefore ultrasound images, which are often noisy, need to be filtered without affecting the key diagnostic details and edges of anatomical structures. In the pre-processing module, we employed two US imaging-specific techniques for enhancing their quality \cite{hussain2017differential}. First, we used fuzzy modeling to distinguish noise from edges and other diagnostic details in the image. We used the fuzzy filter proposed in \cite{nadeem2019fuzzy}, which effectively balances the degree of noise removal and edge preservation. It used the fuzzy logic-based uncertainty modeling, which acquires the statistical parameters (mean and variance) of the local window $\mathbb{W}_l$ in the image to find its similarity with non-local window $\mathbb{W}_{nl}$. Then, the degree of similarity of the similar local and non-local regions is computed using euclidean distance as:
\begin{equation}
    d_s = ||\mathbb{W}_l - \mathbb{W}_{nl}||
\end{equation}
Finally, the noise-free pixel of image $\tilde{p}_i$ is estimated using the fuzzy centroid technique and is given as:
\begin{equation}
    \tilde{p}_i = \frac{1}{\sum_{j=1}^{n} w_k} \sum_{j=1}^{n}(p_j \times w_j) 
\end{equation}
where $n$ is the count for similar local and non-local regions, $p_j$ is the value of the central pixel of window $\mathbb{W}_j$ and $w_j$ is the weight assigned to the most similar pixel, calculated using Gaussian-shaped fuzzy function with zero means $(\mu=0)$ and noise variance $(\sigma^2)$ as:
\begin{equation}
    w_k(d_s; \sigma, \mu) = \exp^{\frac{-(d_s-\mu^2)^2}{2\sigma^2}}
\end{equation}
The fuzzy filtering will enable the network to extract essential quality-related features rather than being distracted by the noise in the US images. Second, we used the sharpening filter, which emphasizes the boundaries of anatomical structures by increasing the contrast between the bright and dark regions of the US image \cite{anand2013sharpening}, thereby enhancing the overall appearance of the image. We used a 4-neighbors laplacian filter: $[[0,-1,0], [-1,5,-1], [0,-1,0]]$, to sharpen the US images. Finally, horizontal flipping is also employed to augment the dataset and prevent the over-fitting of the model on a small US dataset. This will mirror the anatomical structures in the US images, as they can appear on both sides of the transducer cone during scanning.
\\
\\
\textbf{Testing dataset:} 
The dataset consisted of total $1833$ US images. We randomly split the dataset into training and testing set with $90:10$ ratio. We used the labelled testing dataset, as shown in Fig. \ref{fig:dataset}, to validate the proposed framework. The labelling criteria is adopted from the internationally prescribed 5-level quality assessment scale \cite{duan20215g}. Three radiologists with more than $15$ years of experience in abdomen radiology labeled the dataset. The score assesses the image based on the diagnostic quality, which depends on structural shape/size/placement and acceptability of artifacts. We measured inter-rater reliability with Intra-Class Correlation (ICC) \cite{koo2016guideline} and obtained a high value of $0.965$, indicating excellent agreement. However, the supervised US-IQA model in \cite{song2022medical} achieved only $75.3\%$, $77.2\%$, and $76.3\%$ accuracy with respect to three experts, while accuracy for the average rating of three experts is $87.3\%$. These results reveal that inter- and intra-observer perceptual variations pose challenges for supervised US-IQA models. Further, we used Silhouette Coefficient (SC) analysis \cite{zhou2014automatic} on the dataset to choose an optimal value for the number of clusters. For $3$ clusters, SC value was $0.7628$, while for $4$ and $5$ clusters, it was $0.6381$ and $0.5829$, respectively. Therefore, we used three quality labels for the testing set as Below Average ($A-$), Average ($A$) and Above Average ($A+$), where $A$ is collectively used for images rated as $2$ and $3$; and $A+$ for images rated as $4$ and $5$ by experts.
\vspace{-4mm}
\section{Results and Analysis}
\subsection{Implementation details}
The proposed model was implemented using Python $3.8$ and Pytorch $1.11$. The pretraining, training and testing have been conducted using a Dell Workstation with Nvidia GeForce RTX $3090$ GPU having $24$ GB of memory. We used the greyscale images and resized them to $224 \times 224$, and the training set included our pre-processed dataset. The size of window $\mathbb{W}$ used for the fuzzy filter is $5 \times 5$.  The weights of the VAE have been learned in the pretraining phase. The latent space dimension $N$ is kept to $80$ to extract rich features in the latent space for efficient clustering. The choice of latent space dimension has been decided iteratively based on the clustering accuracy scores. The batch size used is $128$ with a learning rate of $0.01$. For the fine-tuning phase, VAE was trained by embedding the clustering module, as described in Section \ref{sec:clustering}. The value of $\gamma$ used in eq. \eqref{loss function} is $0.1$. The batch size used is $64$ and the learning rate is $0.005$. The model was trained for $200$ epochs with early stopping criteria, which is based on the stagnation of the loss function values during training for $10$ epochs.
\subsection{Evaluation metrics}
In order to validate the clustering quality of our model, we used four evaluation metrics, namely Adjusted Random Index (ARI), Adjusted Mutual Information (AMI), Unsupervised Clustering Accuracy (ACC), and  Normalized Mutual Information (NMI) \cite{vinh2010information}. The value of these metrics varies from $0$ to $1$, denoting the accuracy of clustering. 
\subsection{Pre-trained VAE analysis}
We quantitatively analyzed the reconstruction of US images in the test set by the trained VAE using the Structural Similarity Index Measure (SSIM) and Mean Square Error (MSE), as shown in the table \ref{tab:grid}. In order to analyze the effect of pre-processing on extracting rich features from ultrasound images, we also analyzed the reconstructed images without the pre-processed dataset. The image with quality labels A and A+ showed significant improvement in the scores due to the enhancement of anatomical appearance features, thereby enabling the network to focus on key quality features. The image of quality A- has shown slight improvements in scores due to the absence of anatomical structures in the views. 
\begin{table}[ht]
\centering
\caption{Quantitative comparison of testing set to analyze the reconstruction by trained VAE.}
\resizebox{\linewidth}{!}{
\begin{tabular}{ccccccc}
\toprule
\multirow{2}{*}{\textbf{Metric}} & \multicolumn{3}{c}{\textbf{Without pre-processing}} &  \multicolumn{3}{c}{\textbf{With pre-processing}}\\
~ & \textbf{A-} & \textbf{A} & \textbf{A+} & \textbf{A-} & \textbf{A} & \textbf{A+} \\ 
\midrule
SSIM & $0.90$  & $0.85$ & $0.81$ & $0.91$  & $0.93$ & $0.93$\\ 
MSE & $11.01$ & $12.89$  & $13.64$ & $10.34$ & $9.80$  & $9.95$ \\ 
\bottomrule
\end{tabular}}
\label{tab:grid}
\end{table}
\vspace{-0.25cm}
\subsection{Clustering performance analysis}
\begin{figure}[ht]
    \centering
     \includegraphics[trim=0cm 0cm 0cm 0cm,clip,width=0.98\linewidth]{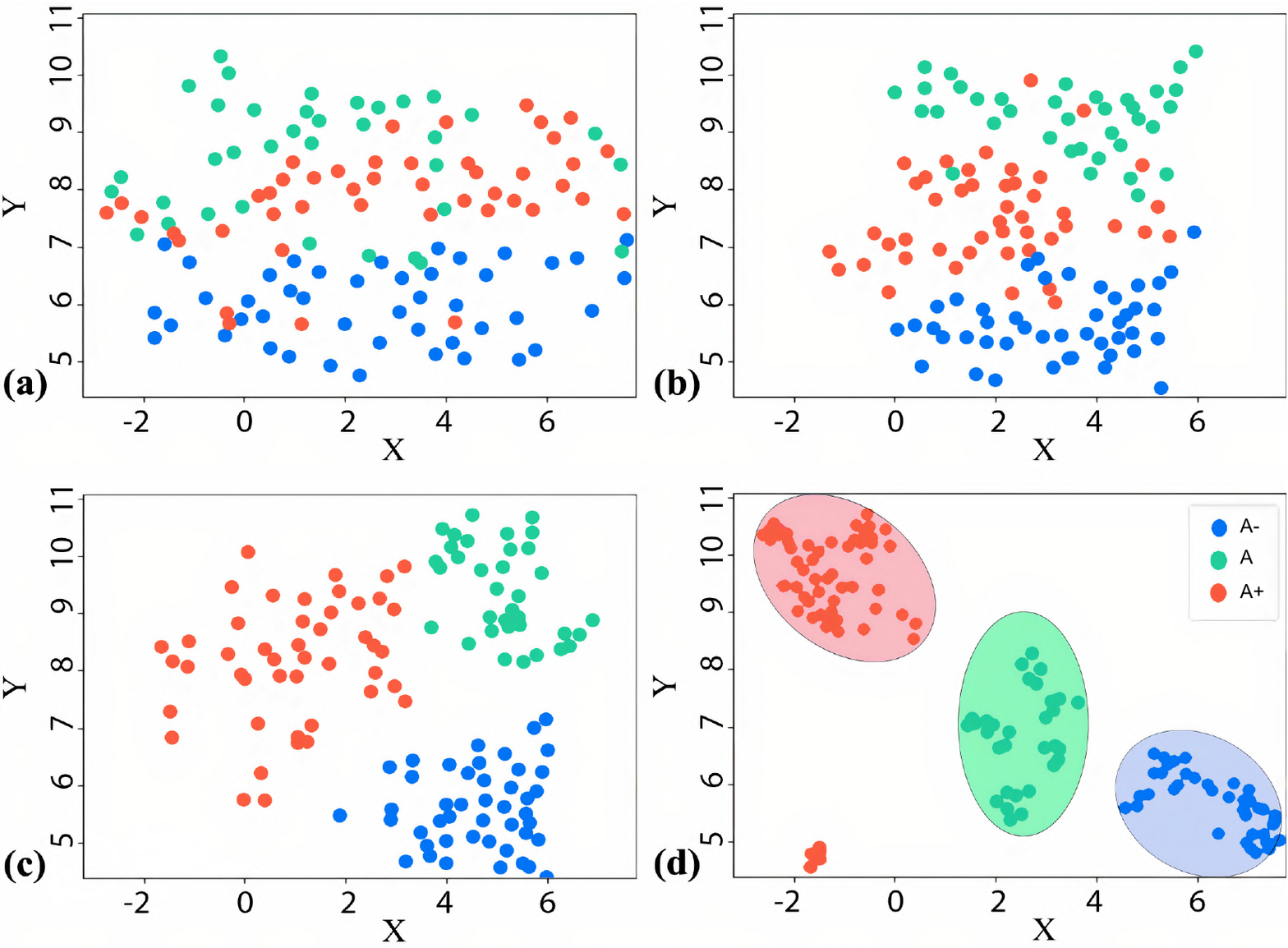}
    \caption{Clustering visualization of US dataset: (a) After pretraining; and after embedding clustering module and training till (b) epoch $70$; (c) epoch $140$; (d) epoch $200$}
    \label{fig:US-TQ}
\end{figure}

\begin{figure}[!ht]
    \begin{subfigure}{0.5\linewidth}
    \centering
    \includegraphics[trim=0cm 0.5cm 0cm 0.6cm,clip,width=\linewidth]{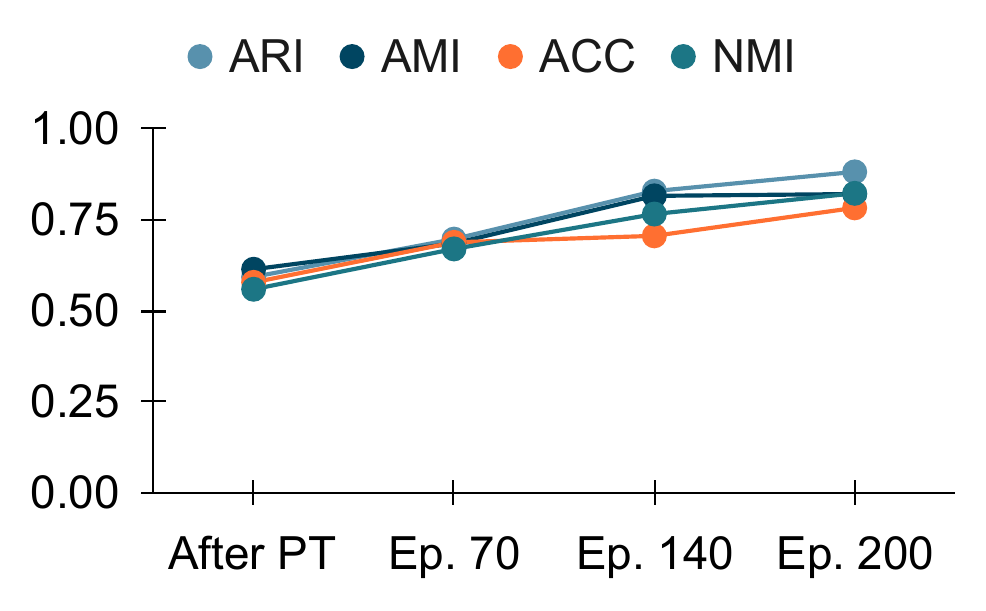}
    \caption{}
    \end{subfigure}%
    \begin{subfigure}{0.5\linewidth}
    \centering
    \includegraphics[trim=0cm 0.5cm 0cm 0.6cm,clip,width=\linewidth]{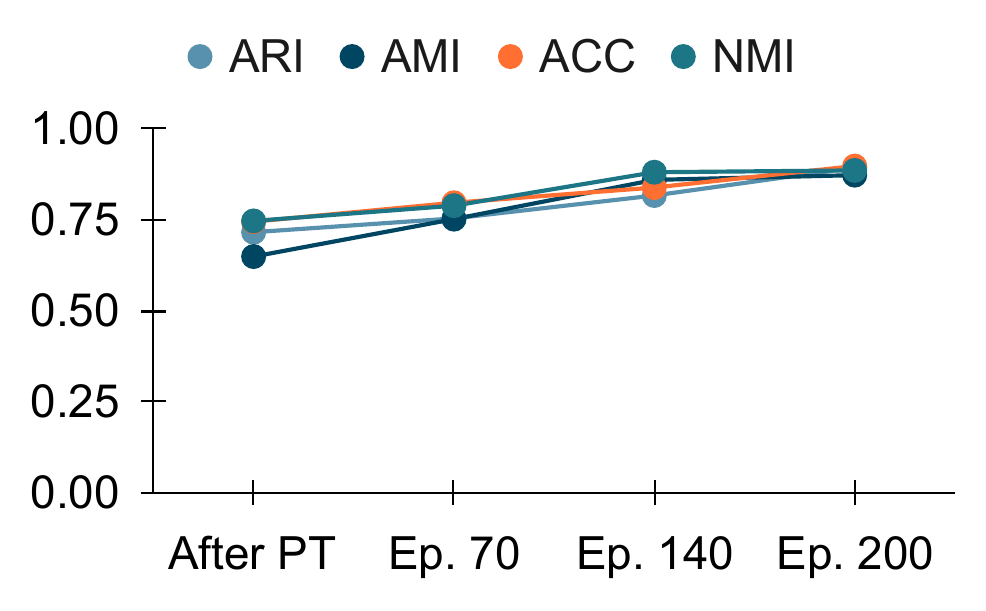}
    \caption{}
    \end{subfigure}%
    \caption{Evaluation metrics comparison for (a) US and (b) MNIST dataset after pre-training (PT) and then training with clustering module.}
    \label{fig:Metrics Comparison}
\end{figure}

\noindent
\textbf{Results with US Image dataset:}
We trained the model on the entire US dataset and visualized the cluster for $183$ test images using UMAP for dimensionality reduction followed by HDBSCAN for label assignment. Fig. \ref{fig:US-TQ} demonstrates that the clusters generated 
after pretraining the VAE has a lot of overlapping points, 
however, the model is progressively getting better after fine-tuning the weights with the clustering module, with the separation of clusters being more evident along with the reduction in overlapping data points. 
This change becomes quantitatively evident, with a $32.8\%$ increase in ARI, $25.2\%$ in AMI, $32\%$ in NMI and $26.2\%$ in ACC scores from the pretraining stage to training with clustering module, as shown in Fig. \ref{fig:Metrics Comparison}(a).
\\
\\
\textbf{Results with MNIST dataset:} We also demonstrated the effectiveness of the proposed model on the MNIST digit dataset \cite{deng2012MNIST}. The values of evaluation matrices are shown in \ref{fig:Metrics Comparison}(b), which demonstrates similar improvements in clustering as observed in the US dataset. Note that ultrasound-specific pre-processing has not been used for MNIST.

\begin{table*}[ht]
\caption{Comparison of clustering performance on US and MNIST dataset using SOTA classical and Deep Embedded Clustering (DEC)-based methods.}
\resizebox{\linewidth}{!}{\begin{tabular}{cllllllllllll}
\toprule
\multicolumn{1}{l}{\multirow{2}{*}{Dataset}} & \multirow{2}{*}{Metrics} & \multirow{2}{*}{$k-$means} & \multirow{2}{*}{t-SNE} & \multirow{2}{*}{UMAP} & \multirow{2}{*}{HDBSCAN} & \multirow{2}{*}{\begin{tabular}[c]{@{}l@{}}PCA\\+$k-$means\end{tabular}} & \multirow{2}{*}{\begin{tabular}[c]{@{}l@{}}DEC-AE\\+$k-$means\end{tabular}} & \multirow{2}{*}{\begin{tabular}[c]{@{}l@{}}DEC-AE\\+t-SNE\end{tabular}} & \multirow{2}{*}{\begin{tabular}[c]{@{}l@{}}DEC-AE\\+UMAP\end{tabular}} & \multirow{2}{*}{{\begin{tabular}[c]{@{}l@{}}DEC-VAE\\+$\boldsymbol{k-}$means\end{tabular}}} & \multirow{2}{*}{{\begin{tabular}[c]{@{}l@{}}DEC-VAE\\+t-SNE\end{tabular}}} & \multirow{2}{*}{\textbf{\begin{tabular}[c]{@{}l@{}}US2QNet\end{tabular}}} \\
\multicolumn{1}{l}{}                         &                          &                         &                                                                        &                       &                         &                        &                                                                        &                                                                       &                                                                      &                                                                                   &                                                                                 &                                                                                 \\ \midrule
\multirow{4}{*}{\textbf{US Images}}        & ARI                      & 0.296                   & 0.385                                                                 & 0.384                 & 0.374                   & 0.366                  & 0.487                                                                  & 0.502                                                                 & 0.547                                                                & 0.532                                                                             & 0.619                                                                           & \textbf{0.882}                                                                           \\
                                             & AMI                      & 0.245                   & 0.394                                                                  & 0.379                 & 0.335                   & 0.389                  & 0.393                                                                  & 0.410                                                                  & 0.446                                                                & 0.545                                                                             & 0.637                                                                           & \textbf{0.821}                                                                           \\
                                             & ACC                      & 0.342                   & 0.389                                                                  & 0.384                 & 0.326                   & 0.360                  & 0.536                                                                  & 0.627                                                                 & 0.673                                                                & 0.587                                                                             & 0.742                                                                           & \textbf{0.783}                                                                         \\
                                             & NMI                      & 0.287                   & 0.371                                                                  & 0.354                 & 0.330                   & 0.296                  & 0.288                                                                  & 0.419                                                                 & 0.453                                                                & 0.553                                                                             & 0.585                                                                           & \textbf{0.823}                                                                          \\ \midrule
\multirow{4}{*}{MNIST}                  & ARI                      & 0.570                   & 0.468                                                                  & 0.386                 & 0.438                   & 0.546                  & 0.596                                                                  & 0.834                                                                 & 0.868                                                                & 0.556                                                                             & 0.887                                                                           & \textbf{0.896}                                                                           \\
                                             & AMI                      & 0.595                   & 0.466                                                                  & 0.547                 & 0.346                   & 0.528                & 0.623                                                                  & 0.694                                                                 & 0.845                                                                & 0.619                                                                             & 0.870                                                                            & \textbf{0.897}                                                                          \\
                                             & ACC                      & 0.546                   & 0.834                                                                  & 0.846                 & 0.569                   & 0.468                  & 0.799                                                                  & 0.857                                                                 & 0.847                                                                & 0.821                                                                             & 0.715                                                                           & \textbf{0.853}                                                                          \\
                                             & NMI                      & 0.502                   & 0.825                                                                  & 0.842                 & 0.532                   & 0.540                  & 0.725                                                                  &  0.704                                                                 & 0.886                                                               & 0.796                                                                             & 0.860                                                                           & \textbf{0.883}                                                                          \\ \bottomrule
\end{tabular}}
\label{tab:Performance}
\end{table*}
\vspace{-0.15cm}
\subsection{Comparison with State-of-the-art methods}
\begin{figure}[ht]
    \centering
     \includegraphics[trim=0cm 0cm 0cm 0cm,clip,width=\linewidth]{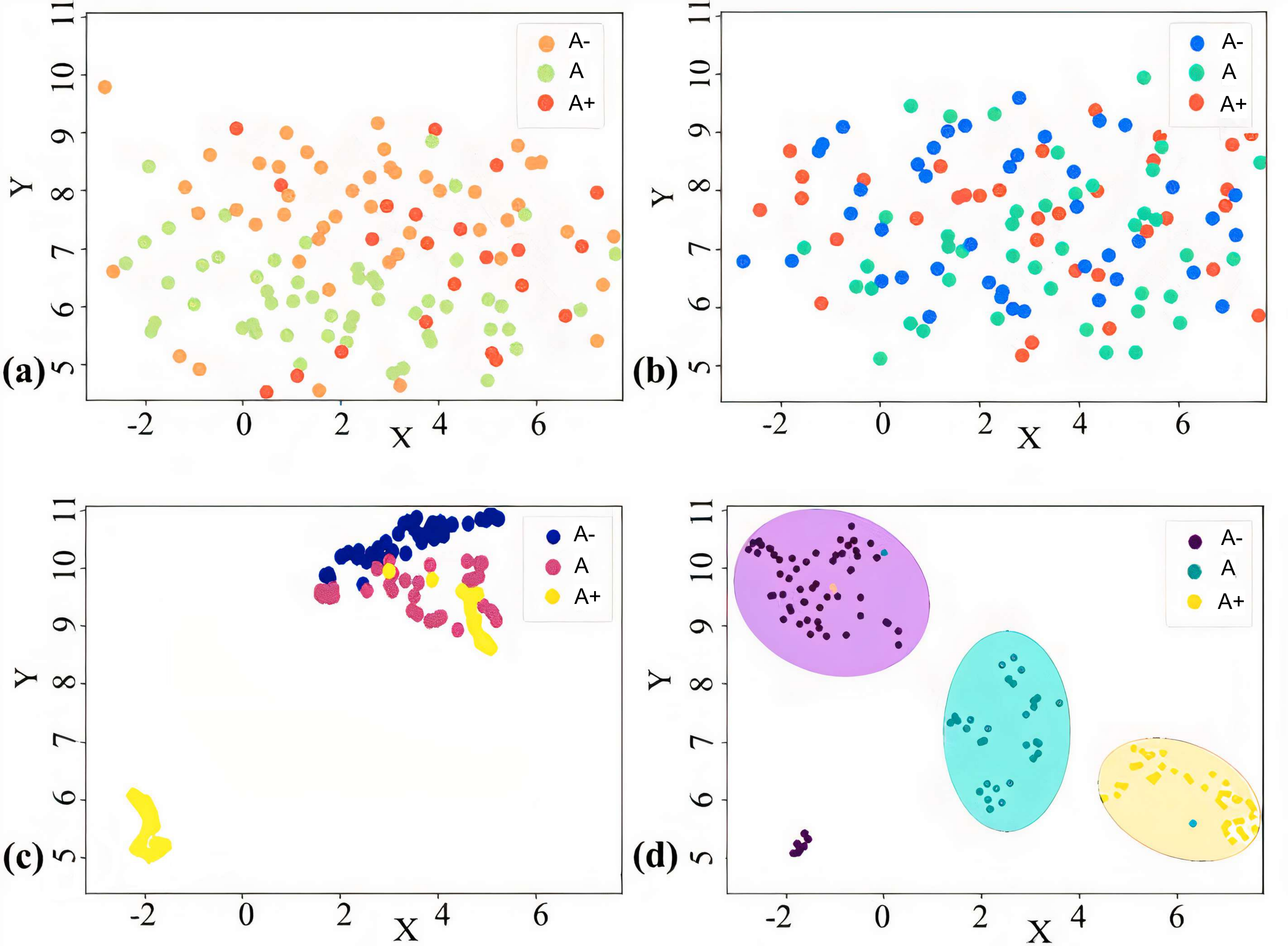}
    \caption{Visualization of clustering obtained using (a) UMAP; (b) t-SNE; (c) DEC-AE+UMAP; (d) DEC-VAE+UMAP (US2QNet) for US dataset.}
    \label{fig:TSNE_CAE_QGC-VAE}
\end{figure}

In this experiment, we compared our method with five classical clustering methods, including the $k-$means, t-SNE, UMAP, HDBSCAN and PCA with $k-$means (PCA+$k-$means) as well as state-of-the-art DEC methods, such as DEC-CAE with $k-$means, t-SNE and UMAP for dimensionality reduction in post-processing. To compare the DEC-VAE-based method with the AE versions, we used similar versions referred to as DEC-VAE+$k-$means, DEC-VAE$+$t-SNE and DEC-VAE$+$UMAP (US2QNet). The results are shown in Table \ref{tab:Performance}. It is noted that our proposed method, US2QNet has obtained the highest performance on US and MNIST dataset. It outperforms classical clustering methods including $k-$means, t-SNE, UMAP, HDBSCAN and  PCA$+k-$means for the US dataset on the NMI score by $65.1\%$, $54.9\%$, $56.9\%$, $59.9\%$ and $64.0\%$, respectively. The DEC-VAE with $k-$means, t-SNE and UMAP reports an increase in NMI score in contrast to its corresponding versions of DEC-AE by $47.9\%$, $28.4\%$ and $44.9\%$, respectively. Similar improvements have also been noticed in ARI, AMI, ACC and NMI on MNIST and US dataset, indicating that the proposed method is effective in unsupervised US-IQA.
The performance of different clustering methods for the US image dataset is visualized in Fig. \ref{fig:TSNE_CAE_QGC-VAE}. The results show that the clusters generated by the proposed method (Fig. \ref{fig:TSNE_CAE_QGC-VAE}(d)) are more evenly distributed and individually closely packed than the other classical and DEC-based approaches. Further, we calculated the ACC values for $4$ and $5$ clusters, resulting in $0.543$ and $0.542$ respectively. From this, it can be asserted that the proposed framework is appropriate for $3$ clusters.
\subsection{Ablation study} To analyze the contribution of different components of the proposed method in enhancing the clustering of US image quality, we performed the ablation study of our model. The results are reported in Table \ref{tab:ablation}. We compared the proposed method with Pre-Processing (PP) only and with Pre-Training (PT) only. The metric used in the plot is ACC. It is evident from the figure that PP and PT modules have contributed to enhancing the performance of the proposed method.
Without PT and PP modules, the ACC value is $0.544$. The model with only pre-processed dataset without pre-training has an ACC score of $0.569$, while using the pre-trained VAE and unprocessed dataset, the model reported an ACC score of $0.643$. However, with the introduction of PP along with PT, the score impressively increased to $0.783$.
\begin{table}[ht]
\centering
\caption{ACC scores for ablated versions of US2QNet with and without pre-processing (PP) and pre-training (PT)}
\resizebox{\linewidth}{!}{
\begin{tabular}{cccc}
\toprule
\textbf{without PT, PP} & \textbf{with PP only} & \textbf{with PT only} & \textbf{with PT and PP} \\ 
\midrule
$0.544$ & $0.569$  & $0.643$ & $0.783$ \\ 
\bottomrule
\end{tabular}}
\label{tab:ablation}
\end{table}
\subsection{Real-time implementation}
\begin{figure}[ht]
    \centering
     \includegraphics[trim=0cm 0cm 0cm 0cm,clip,width=0.7\linewidth]{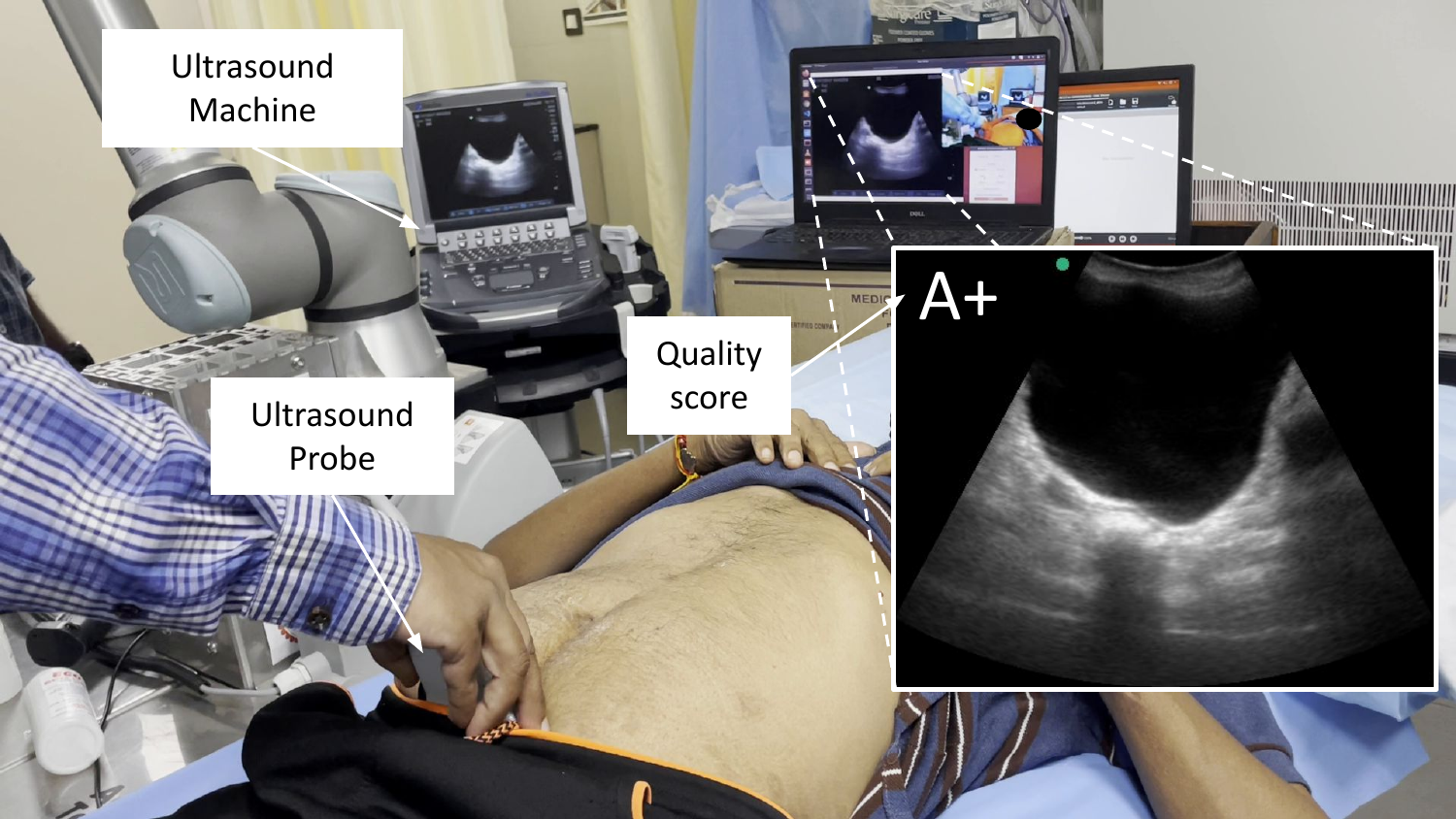}
    \caption{Real-time implementation of US2QNet for assisting the manual ultrasound procedure}
    \label{fig:real_time}
\end{figure}

We have integrated the US2QNet with the ultrasound machine in order to evaluate its real-time efficacy. The US video is captured by the Sonosite M-TURBO ultrasound machine and is transferred to the GPU laptop using Epiphan DVI2USB 3.0 (Epiphan Video, Canada). The manual scanning of the pelvic region is conducted, as shown in Fig. \ref{fig:real_time}.  It has been found that US2QNet efficiently evaluates the quality of US images per frame in $0.0552$ seconds. The sonographer found it to be a useful tool for assisting the novice operator in conducting the procedure, however, they would like the range of quality scoring to be increased as per clinical protocols for precisely guiding the probe maneuvers.
\vspace{-3mm}

\vspace{-0.5cm}
\section{Conclusion and Future work}
We introduced the first-of-its-kind UltraSound Image Quality Assessment (US-IQA) using unsupervised learning. This will alleviate the burden and uncertainty associated with manual annotation of US image quality for supervised learning. The proposed model, US2QNet, leverages the variational autoencoder in conjunction with three modules, pre-processing, clustering and post-processing for effectively enhancing, classifying and visualizing the quality features representations of US images. The validation on urinary bladder US images demonstrated that the proposed framework could generate clusters with $78\%$ accuracy and performs better than state-of-the-art methods. We hope our preliminary results will generate the community's attention towards this efficient but previously ignored topic of unsupervised US-IQA. In future, we would aim for an end-to-end DNN framework without pre-processing. Further, we would incorporate an automatic selection of quality clusters rather than fixing them beforehand. Additionally, our future work will demonstrate the applicability of the proposed framework for assisting autonomous robotic ultrasound \cite{tr-us}. 
\bibliography{references} 
\bibliographystyle{ieeetr}
\end{document}